\documentstyle[12pt]{article}

\def\beq{\begin{equation}}
\def\eeq{\end{equation}}
\def\bea{\begin{eqnarray}}

\def\eea{\end{eqnarray}}
\def\ds{\displaystyle}
\def\sz{\scriptsize}
\def\ni{\noindent}
\def\ul{\underline}
\def\req#1{(\ref{#1})}
\def\ie{{\it i.e.}\ }

\begin{document}

\begin{center}
{\Large\bf Conformal Sector in $D=6$ Quantum Gravity}

\vspace{1cm}

Sergei D. Odintsov \footnote{E-mail: odintsov @ ebubecm1.bitnet.}
and August Romeo, \\
\vspace{0.5cm}
Tomsk Pedagogical Institute, \\          634041 Tomsk, Russian
Federation,\\
and             Department E.C.M., Faculty of Physics,
University of  Barcelona, \\  Diagonal 647, 08028 Barcelona, Spain

\end{center}

\vspace{1cm}

\ni{\bf Abstract.}
We discuss the conformal factor dynamics in $D=6$. Accepting the
proposal that higher-derivative dimensionless terms in the anomaly-induced
effective action may be dropped, we obtain a superrenormalizable (like
in $D=4$) effective theory for the conformal factor.
The one-loop analysis of this theory gives the anomalous scaling
dimension for the conformal factor and provides a natural mechanism to
solve the cosmological constant problem.

\vspace{1cm}

\ni{\large\bf 1. Introduction.}
After the discovery of the conformal anomaly \cite{DDI} (see also \cite{BC})
it was found that it plays an important role in different
physical situations;
among them one can mention string theory \cite{P}, the C-theorem and
its generalizations \cite{Z}, local and non-local anomaly-induced action
in $D=4$
\cite{R,AMPRD}, anomaly structure in $D$ dimensions \cite{DS} and some
other issues \cite{BvN}.

In the recent work \cite{AMPRD} (see \cite{AMM}-\cite{EOS}, \cite{OP}
for further development),
the effective theory for the conformal factor induced by
the conformal anomaly in $D=4$ quantum gravity has been suggested.
The model of ref. \cite{AMPRD} is supposed to describe  the effective theory of
quantum gravity in the infrared region (at large distances) before the GUT
epoch. Note that the model of ref. \cite{AMPRD} was constructed in close
analogy with $D=2$ induced quantum gravity \cite{P} which, however, may
be realized exactly due to specific properties of the dimension $D=2$.

A very interesting question appears: can a reliable effective theory
of quantum gravity based on anomaly-induced dynamics be constructed for
increasing (even) dimensions $D>4$? That is the purpose of this work:
to clarify some issues connected with this question, using $D=6$ as an
explicit example. In particular, we discuss the structure of the
anomaly-induced action and construct the conformal sector in
$D=6$ quantum gravity.

\ni{\large\bf 2. $D$-dimensional theory.}
First of all, let us describe a general situation in a $D$-dimensional
spacetime ($D$ is even). Starting from conformally invariant free matter
in such a space, we may explicitly find the conformal anomaly $T_A$
which results from using regularization to remove the
infinities \cite{DDI}.
$T_A$ consists of $D$-dimensional geometrical invariants which may be divided
into two groups: conformally invariant ones and total derivative
invariants. In the simplest case, $D=2$, $T_A=cR$, where $R$ is a total
derivative and the central charge $c$ consists of the contribution of
different spin fields (including the gravitational field itself).

Choosing the simplest conformal parametrization
\beq g_{\mu \nu}= e^{2 \sigma} \eta_{\mu \nu}, \label{conmet} \eeq
where $\sigma$ is the conformal factor and $\eta_{\mu \nu}$ is the flat metric,
one can present the conformal anomaly as
\beq T_A={\delta \over \delta \sigma(x)} S_{\rm anom} \label{TderS} \eeq
and find the anomaly-induced action $S_{\rm anom}$ by explicit solution of
\req{TderS} \cite{P,R}. Note that only in $D=2$ does parametrization
\req{conmet} completely fix the gauge, and hence $S_{\rm anom}$ is unique.
In $D\neq 2$ $S_{\rm anom}$ is defined only up to some conformal
invariant.

Integrating eq. \req{TderS} in parametrization \req{conmet} one can find the
general structure of $S_{\rm anom}$ in the form
\beq
S_{\rm anom}=\int d^Dx \left\{
\tau \sigma\Box^{D/2}\sigma + V(\sigma, \nabla_{\mu}\sigma, \Box\sigma )
\right\},
\label{SanomI}\eeq
where $\tau$ is some numerical coefficient and $V$ is a complicated
interaction potential whose structure is not unique. In particular,
for $D=2$ $V=0$,
for $D=4$ $V=\tau_1 [ \Box\sigma + (\partial_{\mu}\sigma)^2 ]^2$,
and so on. The coefficient $\tau_1$, as well as $\tau$, are defined by
by the coefficients of the conformal anomaly (and hence by the number of
fields of different spins in the theory).

In order to describe the anomaly-induced dynamics, one has to add to
$S_{\rm anom}$ \req{SanomI} the classical action of the gravitational field
(in parametrization \req{conmet}). In $D=2$ this is simply the cosmological
constant term, and in $D=4$ the Einstein gravity with cosmological term.
Then, in $D=2$ we have the complete theory. However, in $D>2$ we have to
argue (as was done in \cite{AMPRD}) that the contribution from spin-2
excitations is suppressed and we may consider the effective theory of
the conformal factor. This proposal was partially justified in $D=4$
\cite{AMM} by showing that spin-2 gravitational degrees of freedom,
(for a particular model of quantum gravity) in some approximation, only
give a finite contribution to the coefficients
of the conformal anomaly, and hence may be taken into account afterwards.
We will accept that a similar property holds in general $D$. Another
proposal which has been made in \cite{AMPRD} is that $V$ in
\req{SanomI} may be dropped.
As was shown in ref. \cite{AMPRD}, in $D=4$ this may be
justified by the fact that the effective $\sigma$-theory possesses
an infrared fixed point, and setting the corresponding coupling constant
to zero leads to $V=0$ at this fixed point, where we describe
the effective theory for quantum gravity in infrared.
We will again accept
this proposal in arbitrary $D$. Of course, it is extremely difficult
to check in $D=6$, due to the number of higher-derivative dimensionless
terms in the effective action (only one such term is present at $D=4$).

\ni{\large\bf 3. Conformal factor theory.}
Let us start the construction of the trace anomaly-induced action in
the $D=6$ theory. The standard strategy, applied in the $D=4$ theory
is to start from the conformal anomaly, integrate over it,
and obtain the anomaly-induced action. That is the action (plus
Einstein theory) which was used in ref.\cite{AMPRD} to describe the IR
sector of quantum gravity.
Such procedure is not difficult to do in $D=4$, where the conformal anomaly
for conformally invariant fields includes only 3 terms.

However, in $D=6$ life is much harder. In particular, let us consider
the theory for a free conformally invariant scalar field in $D=6$:
\beq
S=\int d^6x \sqrt{-g} \left\{
-{1 \over 2} \varphi\Box\varphi + {1 \over 10} R \varphi^2
\right\} .
\eeq
Then, the conformal anomaly (without total derivative terms like
$\Box R$) is given by
\beq
\begin{array}{ll}
\ds T_A={1 \over (4 \pi)^3} \int d^6x \sqrt{-g}&\ds \left\{
-{23 \over 12 \cdot 9450} R \Box R
+ { 13 \over 22680 } R_{\mu \nu} \Box R^{\mu \nu}
+{1 \over 6480} R^{\mu \nu \alpha \beta} \Box R_{\mu \nu \alpha \beta}
\right. \\
&\ds -{1 \over 6}{1 \over 27000}R^3
+{1 \over 180 \cdot 30} R R_{\mu \nu} R^{\mu \nu}
-{1 \over 180 \cdot 30}
R R_{\mu \nu \alpha \beta} R^{\mu \nu \alpha \beta} \\
&\ds -{1 \over 1260} R_{\mu \nu} R^{\mu}_{\alpha} R^{\nu \alpha}
+{1 \over 2268} R_{\mu \nu} R_{\alpha \beta} R^{\mu \nu \alpha \beta}
-{1 \over 5670}
R_{\mu \nu} R^{\mu \lambda \rho \sigma} R^{\nu}_{\lambda \rho \sigma}
\\
&\ds \left. +{1 \over 1890} R_{\mu \nu \rho \sigma}
R^{\mu \nu}_{\alpha \beta} R^{\rho \sigma \alpha \beta}
\right\}
\end{array}
\label{longS}\eeq

Taking into account all the total derivative terms (by using the
corresponding $a_3$-coefficient of the Schwinger-De Witt
expansion\cite{DW},
see \cite{Gi}), the number of terms in \req{longS} increases a
lot. The straightforward integration of such an expression over the
conformal anomaly is a very hard task.

Here we shall outline an alternative and efficient method to construct
the required action, already proposed in \cite{AMPRD}. The chief idea is
that, according to the observations made in that paper, both scale and
conformal invariance must be preserved (taking into account the
transformation of the integration measure $d^Dx$).
This fixes ---up to total divergence--- not just the type
of terms which can be present, but also their relative coefficients.
Let $\varphi=e^{\sigma}$;
in order to maintain scale invariance in $D=6$, we must have a
combination of (integrated) quotients containing the same
number of $\varphi$'s in the numerator and in the denominator and
with six derivatives in the numerator:
\beq
\begin{array}{lll}
\ds
\left( \partial\varphi \over \varphi \right)^2
\left( \partial\varphi \over \varphi \right)^2
\left( \partial\varphi \over \varphi \right)^2&&\\
\ds
\left( \partial\varphi \over \varphi \right)^2
\left( \Box\varphi \over \varphi \right)^2,
&\dots,
&\ds
{ \partial_{\mu}\varphi \partial_{\nu}\varphi \partial_{\lambda }\varphi
\partial^{\mu} \partial^{\nu} \partial^{\lambda} \varphi
\over
\varphi^4}, \\
\ds
\left( \Box\varphi \over \varphi \right)^3,
&\dots,
&\ds
{ \partial_{\mu}\varphi
  \partial_{\nu} \partial_{\lambda }\varphi
 \partial^{\mu}  \partial^{\nu} \partial^{\lambda}\varphi
\over
\varphi^3}, \\
\ds
{ \partial_{\mu}\Box\varphi  \partial^{\mu}\Box\varphi
\over \varphi^2 },
&\dots,
&\ds
{ \partial_{\mu}\varphi  \partial^{\mu}\Box^2 \varphi
\over \varphi^2 }, \\
\ds
{ \Box^3 \varphi \over \varphi }&&
\end{array}
\eeq
On the whole, there are 23 of them.
Taken as a set, they are not conformally invariant.
In fact, the most general
Lagrangian will be a sum of all the independent conformally-invariant
combinations which can be constructed with these building
blocks.
We put the infinitesimal expression of a conformal transformantion in
the way:
\beq
\begin{array}{lll}
\delta\varphi(x)&=&2 \tilde\alpha \varepsilon \cdot x \varphi(x), \\
\delta x^{\mu}&=&-\varepsilon_{\nu}
(2 x^{\mu} x^{\nu} - \eta^{\mu \nu} x^2),
\end{array}
\eeq
where the $\varepsilon_{\mu}$'s are the infinitesimal parameters
and $\tilde\alpha=D/2-1$, as usual.

The quantity
$
d^Dx \left( \Box\varphi \over \varphi \right)^3
$
is already conformally invariant in $D=6$ and will therefore be one of
the pieces in our Lagrangian, as it stands. However, this case is rather
execeptional, and in general one has to perform lengthy examinations
of the transformed versions for each term so as to construct the desired
invariant combinations. We illustrate this with an example: the terms
$
d^Dx \left( \Box\varphi \over \varphi \right)^2
\left( \partial\varphi \over \varphi \right)^2,
$
$
d^Dx
\left( \partial_{\mu} \partial_{\nu}\varphi \over \varphi \right)^2
{ \Box\varphi \over \varphi }
$
are not conformally invariant when taken separately. However, the linear
combination
$
d^Dx \left[ 5 \left( \Box\varphi \over \varphi \right)^2
\left( \partial\varphi \over \varphi \right)^2
-2 \left( \partial_{\mu} \partial_{\nu}\varphi \over \varphi \right)^2
{ \Box\varphi \over \varphi } \right]
$
actually is (in $D=6$), and may be included in our action. Thus,
\beq
S_{\mbox{\sz anom}}= \int d^Dx \left\{
\tau^2 \left( \Box\varphi \over \varphi \right)^3
+\rho
\left[ 5 \left( \Box\varphi \over \varphi \right)^2
\left( \partial\varphi \over \varphi \right)^2
-2 \left( \partial_{\mu} \partial_{\nu}\varphi \over \varphi \right)^2
{ \Box\varphi \over \varphi }
\right]
+\dots
\right\}
\label{Sconf}\eeq
We will not write explicitly all terms in \req{Sconf}, as the
corresponding expression would take more than one page and is not really
used in the explicit analysis below.

After taking this process to the end, one should get an action
$S_{\mbox{\sz ext}}$ giving the anomaly-induced action with contributions
from the classical theory, \ie
$S_{\mbox{\sz ext}}=S_{\mbox{\sz anom}}+S_{\mbox{\sz grav}}$.
                           Of course, the connection between the
coefficients in such an action and those in the conformal anomaly is
absent.

In $S_{\mbox{\sz ext}}$,
the first term $S_{\mbox{\sz anom}}$ corresponds to the integration of
the conformal anomaly, while the second comes from the corresponding
ones in the classical action for gravity in $D=6$:
\beq
\ds S_{\mbox{\sz grav}}=\int d^6x \sqrt{-g}
\left\{
\lambda + \tilde\gamma_1 R
\right\} .
\label{Sgrav}
\eeq

Thus, our analog   of the $4D$
action in ref. \cite{AMPRD} looks like (we drop also derivative
interactions in the dimensionless term $S_{\mbox{\sz anom}}$, supposing that,
like in $D=4$,
this corresponds to an IR stable point of the theory or there exists
some other mechanism for its vanishing.)


\[  S_{\mbox{\sz ext}}= \int d^Dx {\cal L}[\sigma(x)], \]
\beq
\begin{array}{ll}
\ds {\cal L}[ \sigma ]=&\tau^2 \sigma \Box^3 \sigma
+ \lambda e^{6 \sigma}
+ \gamma_1 e^{4 \sigma} [\Box\sigma + 2(\partial\sigma)^2] .
\end{array}
\eeq
This expression is taken to be written in Euclidean space. Then,
(we adopt new notations from this point), changing $\sigma$ as
\beq \sigma \to \alpha\sigma, \eeq
and ---after dividing by $\alpha^2$--- the Lagrangian for the
conformal-anomaly induced dynamics can be put into the form
\beq
\ds {\cal L}[ \sigma ]=
\ds \tau^2 \sigma \Box^3 \sigma
+{\lambda \over \alpha^2} e^{6 \alpha \sigma}
-2\gamma_1 e^{4 \alpha \sigma}(\partial\sigma)^2 .
\label{LPhi}\eeq
That will be our starting point. As we will see, this theory in $D=6$
is superrenormalizable, as in $D=4$.

\ni\ul{Feynman rules.}
The Feynman rules for the free propagator and tree-level vertices
derived from this action have the following expression

\ni{\bf Free propagator}

\beq
-{1 \over 2}
{1 \over \tau^2 k^6 +2\gamma_1 k^2 -18 \lambda }
\eeq

\ni{\bf $\lambda$-vertex}

\beq -{\lambda\over\alpha }(6 \alpha)^n \eeq

\ni{\bf $\gamma_1$-vertex}

\beq
\ds-{\gamma_1 \over 4 \alpha^2}(4 \alpha)^n
\sum_{j,l \atop 1 \le j < l \le n} (p_j \cdot p_l)
\eeq

\ni\ul{Regularization and renormalization.}
For a general graph, let $L$ denote the number of loops, $I$ the number
of internal propagators, and $V_{\gamma_1}$ the
number of $\gamma_1$-vertices.
Combining elementary power counting and the usual topological relation,
we derive the value of the superficial degree of divergence ${\cal D}$:
\beq
\left.
\begin{array}{c}
{\cal D}=6L-6I+2V_{\gamma_1} \\
L=I-V+1, \hspace{1cm} V\equiv V_{\lambda}+V_{\gamma_1}
\end{array}
\right\}
\Longrightarrow
{\cal D}=6-6V_{\lambda}-4V_{\gamma_1}.
\eeq
Thus, there are only two possible classes of superficially divergent
diagrams, which are typified by
\beq
\begin{array}{lll}
V_{\lambda}=1&V_{\gamma_1}=0&{\cal D}=0, \\
V_{\lambda}=0&V_{\gamma_1}=1&{\cal D}=2.
\end{array}
\eeq
Note that, as a result of working in $D=6$, this theory is even more
convergent that the analogous one developed for $D=4$ in \cite{AMPRD}.

For $L=1$ (one-loop), the calculation of any of these diagrams
involves ---at least for the simplest combinations of vanishing
external momenta--- integrals of the generic sort
\beq
{\cal I}_I^p\equiv \int{d^Dk \over (2\pi)^D}
{k^{2p} \over (k^6-ak^4-bk^2-c)^I} .
\eeq

By dimensional regularization in
\beq D=6-2\varepsilon  \eeq
dimensions, we have been able to calculate the following cases
\beq
\begin{array}{lll}
{\cal I}_I^{3I-3}&=&\ds{1 \over (4\pi)^3} {1 \over 2\varepsilon} +
O(\varepsilon^0), \\
{\cal I}_I^{3I-2}&=&\ds{1 \over (4\pi)^3} {Ia \over 2\varepsilon} +
O(\varepsilon^0), \\
{\cal I}_I^{3I-1}&=&\ds{1 \over (4\pi)^3}
{ I\left[ b+(I+1){a^2 \over 2} \right] \over 2\varepsilon} +
O(\varepsilon^0). \\
\end{array}
\eeq
Since in our theory we have no $k^4$ term,
the second integral is actually finite.
{}From these, it is also possible to obtain integrals like
\beq
{\cal I}_1^{\mu \nu} \equiv \int{d^Dk \over (2\pi)^D}
{ k^{\mu} k^{\nu} \over k^6-ak^4-bk^2-c }
= {1 \over D} {\cal I}_1^1 \delta^{\mu \nu}.
\eeq

All of them will be necessary for the evaluation of the one-loop graphs
to be studied.

\ni\ul{Results for some one-loop diagrams}

\begin{itemize}

\item{\bf $L=1, V_{\lambda}=1, V_{\gamma_1}=0$},
with $n$ external legs carrying no momentum:

\beq
{ \lambda \over 2 \alpha^2 \tau^2 } (6\alpha)^{n+2}
{1 \over (4\pi)^3} {1 \over 2 \varepsilon}
+O(\varepsilon^0)
\eeq

\item
\begin{itemize}

\item{\bf $L=1, V_{\lambda}=0, V_{\gamma_1}=1$},
with $n$ external legs carrying no momentum (derivatives acting
on the loop line only): in this situation, the diagram turns out to be
finite.

\item{\bf $L=1, V_{\lambda}=0, V_{\gamma_1}=1$},
with $n+2$ external legs, two of them carrying momenta $p$ and $-p$,
and the remainig $n$ carrying no momentum (derivatives acting on these
external lines only):

\beq
-{ \gamma_1 \over 8 \alpha^2 \tau^2} (4 \alpha)^{n+4}
{1 \over (4 \pi)^3 }
p^2 {1 \over 2 \varepsilon}
+O(\varepsilon^0)
\label{L1010pmp}\eeq

\end{itemize}

\end{itemize}

\ni\ul{Beta functions.}
By selecting particular cases of the above
diagrams,
we renormalize the two-point function. After taking into account the
classical scaling contributions like in \cite{AMPRD}, we find
the one-loop beta functions for the $\lambda$ and $\gamma_1$
couplings to be
\beq
\begin{array}{llll}
\beta_{\lambda}&=&\ds (6-6\alpha) \lambda_r
&\ds+9{\alpha^2 \over (4 \pi)^3}
{\lambda_r \over \tau^2}
+ \dots, \\
\beta_{\gamma_1}&=&\ds (4-4\alpha) \gamma_{1r}
&\ds-8{\alpha^2 \over (4 \pi)^3}
{\gamma_{1r} \over \tau^2}
+ \dots \\
\end{array}
\label{betas}
\eeq
Similarly, one can find the two-loop,             etc corrections. We
stop at the one-loop stage.
The solutions of $\beta_{\gamma_1}=0$ for $\gamma_{1r}\neq 0$ are
$\ds
\alpha=\alpha_{\pm}\equiv
{ -1\pm \sqrt{1+8/((4\pi)^3\tau^2)} \over 4/((4\pi)^3\tau^2) },
$
This result is analogous to the anomalous scaling dimension
in $D=2$ \cite{PKZ}  or in $D=4$ \cite{AMPRD}. Note that, in the
above relation, $\tau^2$ plays the role of the central charge
in $D=6$ and $\ds\tau_{cr}^2=-8(4 \pi)^{-3}$ corresponds to the
$c$-barrier in $D=6$.

As happens in $D=4$, the classical scaling dimension is obtained from
the positive branch of $\alpha_{\pm}$ in the limit of no QG
($\tau^2 \to \infty$).
\beq
\alpha=\alpha_{+}\equiv
{ -1+ \sqrt{1+8/((4\pi)^3\tau^2)} \over 4/((4\pi)^3\tau^2) },
\eeq
which is associated to the physical solution.

Now, we turn to the analysis of the first relation in \req{betas}. In
$D=4$ dimensions, the analog of this equation has led to fixing the
cosmological constant in terms of the Newton constant \cite{AMPRD}.
Now, in $D=6$ this is no longer the case. Instead, after
substituting the value of $\alpha_{\pm}$ into \req{betas}, we get the
relation
\beq
\beta_{\lambda}=
-{3 \alpha^2 \over (4 \pi)^3}{\lambda_r \over \tau^2} .
\eeq
As one can see, it is not possible to make $\beta_{\lambda}$ vanish
unless $\lambda_r=0$. Hence, one may naturally obtain the solution
of the cosmological constant problem in $D=6$.

Note that, in principle, the anomalous scaling
dimension could be determined from the first expression in \req{betas}.
Then, it would give a different value for $\alpha$, making the
gravitational coupling constant vanish. Eventually, that case should
correspond to the unphysical region of QG, and we do not discuss it.

Summing up, we have constructed the conformal sector of
$6D$ quantum gravity and calculated the anomalous scaling dimension
of the conformal factor. As a result, we have got a very natural
solution of the cosmological constant problem in $D=6$. Of course, there
are still many questions left to be understood in the future. In
particular, our proposal to drop higher-derivative dimensionless terms
---whose number is quite large--- as was done in $D=4$ (where it
happens to correspond to a stable IR fixed point) should be examined
in more detail. Note that this amounts to extremely lengthy
calculations.

Another remark is connected with the fact that in $D=6$, unlike in $D=4$,
one can also add four-derivative terms to the classical gravitational
action. For instance, if we add a $\gamma_2 R^2$ term to \req{Sgrav},
the above one-loop analysis would be completely changed with the result
(instead of \req{betas}):
\beq
\begin{array}{llll}
\beta_{\lambda}&=&\ds (6-6\alpha) \lambda_r
&\ds+{\alpha^2 \over (4 \pi)^3}
\left[ 9 {\lambda_r \over \tau^2}
-{1 \over 6} { \gamma_{1r} \gamma_{2r} \over \tau^4 }
+{3 \over 72} { \gamma_{2r}^3 \over \tau^6} \right] + \dots, \\
\beta_{\gamma_1}&=&\ds (4-4\alpha) \gamma_{1r}
&\ds+{\alpha^2 \over (4 \pi)^3}
\left[
-8{\gamma_{1r} \over \tau^2}
+{61 \over 64}{\gamma_{2r}^2 \over \tau^4}
\right]
+ \dots, \\
\beta_{\gamma_2}&=&\ds (2-2\alpha) \gamma_{2r}
&\ds+{\alpha^2 \over (4 \pi)^3} {1 \over 2}{\gamma_{2r} \over \tau^2}
+ \dots
\end{array}
\eeq
The analysis of these equations is more similar to the case of
ref. \cite{AMPRD}.

Our main conclusion is that the principal possibility of realizing
the conformal factor dynamics in $D>4$ still exists. However, the explicit
details of such a realization are much more complicated.

We would like to thank Ignatios Antoniadis for very stimulating
 discussions and participartion at the early stages of this work.

\end{document}